\documentclass[letter]{IEEEtran}
\IEEEoverridecommandlockouts
\usepackage{cite}
\usepackage{amsmath,amssymb,amsfonts}
\usepackage{algorithmic}
\usepackage{graphicx}
\usepackage{subcaption}
\usepackage{textcomp}
\usepackage{xcolor}
\usepackage[export]{adjustbox}

\usepackage[hyphens]{url}
\usepackage{hyperref}
\usepackage[hyphenbreaks]{breakurl}

\def\BibTeX{{\rm B\kern-.05em{\sc i\kern-.025em b}\kern-.08em
    T\kern-.1667em\lower.7ex\hbox{E}\kern-.125emX}}
\begin{document}

\title{Network Slicing for eMBB and mMTC with NOMA and Space Diversity Reception}

\author{Eduardo Noboro Tominaga\IEEEauthorrefmark{1}, Hirley Alves\IEEEauthorrefmark{1}, Onel Luis Alcaraz López\IEEEauthorrefmark{1}, Richard Demo Souza\IEEEauthorrefmark{2}, Jo\~{a}o Luiz Rebelatto\IEEEauthorrefmark{3}, Matti Latva-aho\IEEEauthorrefmark{1}\\

	\IEEEauthorblockA{
		\IEEEauthorrefmark{1}6G Flagship, Centre for Wireless Communications (CWC), University of Oulu, Finland\\
		\{eduardo.noborotominaga, hirley.alves, onel.alcarazlopez matti.latva-aho\}@oulu.fi\\
		\IEEEauthorrefmark{2}Federal University of Santa Catarina (UFSC), Florian\'{o}polis, Brazil, richard.demo@ufsc.br\\
		\IEEEauthorrefmark{3}Federal University of Technology - Paran\'{a} (UTFPR), Curitiba, Brazil, jlrebelatto@utfpr.edu.br
	}
}

\maketitle

\begin{abstract}
In this work we study the coexistence in the same Radio Access Network (RAN) of two generic services present in the Fifth Generation (5G) of wireless communication systems: enhanced Mobile BroadBand (eMBB) and massive Machine-Type Communications (mMTC). eMBB services are requested for applications that demand extremely high data rates and moderate requirements on latency and reliability, whereas mMTC enables applications for connecting a massive number of low-power and low-complexity devices. The coexistence of both services is enabled by means of network slicing and Non-Orthogonal Multiple Access (NOMA) with Successive Interference Cancellation (SIC) decoding. Under the orthogonal slicing, the radio resources are exclusively allocated to each service, while in the non-orthogonal slicing the traffics from both services overlap in the same radio resources. We evaluate the uplink performance of both services in a scenario with a multi-antenna Base Station (BS). Our simulation results show that the performance gains obtained through multiple receive antennas are more accentuated for the non-orthogonal slicing than for the orthogonal allocation of resources, such that the non-orthogonal slicing outperforms its orthogonal counterpart in terms of achievable data rates or number of connected devices as the number of receive antennas increases.
\end{abstract}

\begin{IEEEkeywords}
5G, network slicing, eMBB, mMTC, NOMA, SIC, Spatial Diversity.
\end{IEEEkeywords}

\section{Introduction}

\par The Fifth Generation (5G) of wireless communications systems features three generic services \cite{ericsson2017}: enhanced Mobile BroadBand (eMBB), Ultra-Reliable and Low Latency Communications (URLLC) and massive Machine-Type Communications (mMTC). The objective of eMBB is to provide connectivity with extremely high peak data rates and relatively low latency, while at the same time guaranteeing moderate and uniform connectivity over the whole coverage area. URLLC is envisioned for real-time monitoring and control applications with extremely demanding requirements in terms of latency and reliability. Finally, mMTC aims at providing ubiquitous connectivity for hundreds or thousands of devices per square kilometer of coverage that feature relatively low software/hardware complexity, low-energy operation, small payloads, low activation probability and relatively high delay tolerance \cite{ericsson2017}. 

\par Despite 5G is still under standardization and initial phases of deployment around the world, the research community has already started working on the definition of the Key Performance Indicators (KPIs) and development of technical solutions for the Sixth Generation (6G) of wireless communication systems. Regarding the mMTC use cases, it is predicted that the number of connected devices will increase substantially, up to hundreds of devices per cubic meter, which poses very stringent requirements on spatial spectral efficiency and required frequency bands for connectivity \cite{6G_White_Paper}. One of the use cases for mMTC towards 6G are the connected industries, that is, the evolution from Industry 4.0 to Industry 5.0. The massive connectivity in industrial setups will enable data-driven solutions for unprecedented levels of personalization and customization of products, as well as the improvement of the operation and performance efficiency \cite{MTC_White_Paper}.

\par Previous generations of wireless communication systems relied mostly on Orthogonal Multiple Access (OMA) schemes. In such schemes, resources that are orthogonal in time and/or frequency are exclusively assigned to different users such that ideally there is no interference among them. However, the major drawback of OMA is that the number of connected devices is limited by the number of available orthogonal radio resources. Meanwhile, Non-Orthogonal Multiple Access (NOMA) techniques have been proposed for the uplink in 5G. NOMA constitues a promising solution to enhance the spectral efficiency and to allow the massive connectivity of users required by mMTC applications. Under NOMA, different users can share the same time/frequency resource through different power allocations or different code signatures. In the case of users with different power levels, the overlapping signals are decoded using Successive Interference Cancellation (SIC) \cite{lien2017}.

\par Another important technique used to enhance the performance of 5G systems is multiple antennas. Since the separation necessary to ensure independence between antennas decreases with the carrier frequency, for higher-frequencies, e.g. mmWave band, a massive number of antennas may be available, which increases the capability for beamforming. On the other hand, for the lower frequency bands, the number of antennas is typically low to moderate, e.g. up to 32 active antennas \cite{zaid2017}. Nonetheless, the available bandwidth in the lower frequency bands is scarce, which may require the combination of multi-antenna techniques with other solutions to increase the number of connected users and the spectral efficiency, e.g. NOMA.


\par Aiming at allowing the three generic 5G services to coexist in the same Radio Access Network (RAN) when the number of radio resources is limited, Popovski \textit{et. al.} \cite{popovski2018} proposed an information-theoretic framework for the slicing of radio resources based on the utilization of NOMA techniques. 
Regarding the specific case of network slicing between eMBB and mMTC, the authors in \cite{popovski2018} proposed a framework where an eMBB device and multiple MTC devices share the same radio resource. In the case of orthogonal slicing, they coexist in a Time Division Multiple Access (TDMA) manner. Conversely, the traffic from both services overlap in the radio resource under the non-orthogonal slicing. In both cases, multiple MTC devices are allowed to transmit concurrently by means of NOMA, and the BS performs SIC  to recover the multiple overlapping signals. However, they did not consider the use of multiple receive antennas on the BS.
Other works also studied the coexistence between eMBB and mMTC, e.g. \cite{vikhrova2019}, \cite{nurul2016} and \cite{kamel2020}, but being restricted to single antenna receivers as well.

\par The uplink scenario where multiple MTC devices are allowed to communicate with one or multiple receivers using NOMA schemes has also been studied on other works. In \cite{onel2018}, the authors studied the uplink mMTC in a large-scale cellular network overlaid with data aggregators  using an analytical framework based on stochastic geometry. In \cite{mo2018}, the authors studied a multi-cell scenario with single cell BSs for a Ultra-Narrow Band (UNB) Low Power Wide Area Network (LPWAN). They considered two different SIC mechanisms: SIC performed locally at each BS without information exchange between BSs, and SIC performed across multiple BSs where BSs can send decoded packets to neighboring cells. The performance of a Long Range (LoRa) network with multiple LoRa devices connected in the uplink with a single antenna BS is studied in \cite{jean2020}. Therein, the BS is allowed to perform a SIC decoding with one iteration to avoid packet losses due to collisions.

\par In the scope of multiple antenna receivers, Liu \textit{et. al. }\cite{liu2016} studied the performance of a single-cell large scale Multi-User-MIMO (MU-MIMO) uplink system. They investigated the performance in terms of outage probability of three linear receivers: Maximum Ratio Combining (MRC), Zero-Forcing (ZF) and Minimum Mean Square Error (MMSE). In their model, all users were allowed to communicate with the BS simultaneously in the same time-frequency resource. However, they did not consider the coexistence of devices with heterogeneous performance requirements.

\par Inspired on the recent works that study MTC uplink scenarios with receive diversity, and specially the setup from \cite{liu2016} and the framework from \cite{popovski2018}, in this work we study the performance of orthogonal and non-orthogonal network slicing in a single-cell scenario where one eMBB device and multiple MTC devices communicate in the uplink with a multi-antenna BS. The BS utilizes an iterative MRC-SIC receiver to decode the multiple packets that arrive simultaneously. The BS utilizes a MRC receiver based on the assumption that the number of MTC devices may be much larger than the number of receive antennas in mMTC scenarios. Differently from \cite{liu2016}, we implement a setup with heterogeneous devices communicating in the uplink. Besides, while in \cite{popovski2018} the authors considered only a single-antenna BS, we evaluate the performance gains provided by multiple receive antennas operating under MRC. The performance is evaluated in terms of achievable data rates and number of connected MTC devices for given reliability requirements of both services. We show, through Monte Carlo simulations, that despite the space diversity reception improves the performance of both slicing schemes, the performance gains are more accentuated for the non-orthogonal slicing, which makes it a more attractive choice than the orthogonal slicing when the BS is equipped with multiple antennas. Given a number of connected MTC devices, the advantage of non-orthogonal slicing over its orthogonal counterpart increases as the number of receive antennas increases. Conversely, non-orthogonal slicing allows us to improve significantly the number of MTC devices that can be connected to the BS as the number of receive antennas increases, for a given target mMTC data rate.

\par The remainder of this paper is organized as follows. In Section \ref{system model} we present the system model and the performance analysis of eMBB and mMTC when the services are considered in isolation. In Section \ref{Slicing between eMBB and mMTC} we formulate the orthogonal and non-orthogonal slicing of radio resources between eMBB and mMTC. The Monte Carlo simulation results of both slicing schemes and numerical discussions are presented in Section \ref{numerical illustration}. Finally, we draw our conclusions in Section \ref{conclusions}.

\section{System Model}
\label{system model}

\par
\begin{figure}[t]
    \centering
    \includegraphics[scale=0.4]{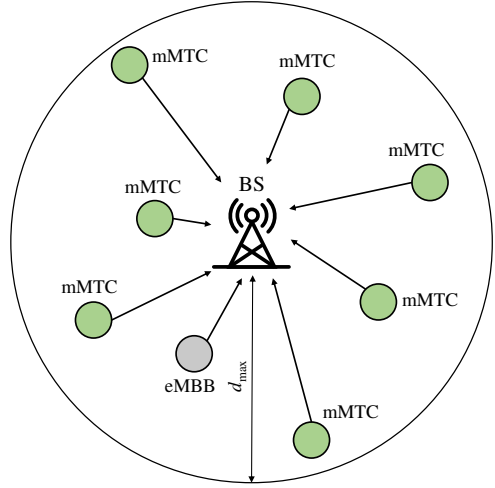}
    \caption{The uplink of a 5G network where an eMBB and multiple MTC devices are connected to a common BS.}
    \label{scenario}
\end{figure}

 We aim at analyzing the uplink performance of a 5G network where a single eMBB device and multiple MTC devices transmit independent packets to a common BS, as illustrated in Fig. \ref{scenario}. The eMBB device and each MTC device are single antenna devices, whereas the BS is equipped with $L$ receive antennas, indexed by $l \in \left\{1,\ldots,L\right\}$.

\par
\begin{figure}[t]
    \centering
    \includegraphics[scale=0.4]{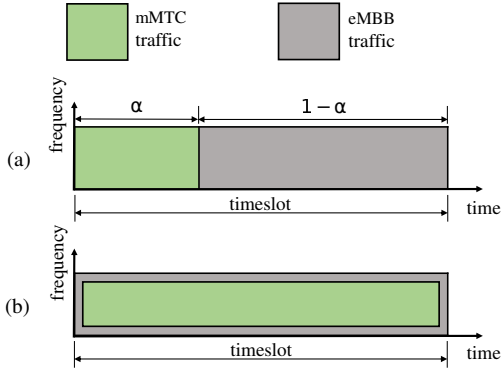}
    \caption{Orthogonal (a) and non-orthogonal (b) slicing of radio resources between eMBB and mMTC.}
    \label{slicing}
\end{figure}

 We consider a time-frequency resource composed of one timeslot $t$ in a single frequency channel $f$ that can be shared by the eMBB and $M$ MTC devices either under orthogonal or non-orthogonal slicing schemes, as illustrated in Fig. \ref{slicing}. Under the orthogonal slicing, a fraction $\alpha$ of the timeslot is allocated exclusively to the mMTC traffic, while the remaining of the timeslot is allocated exclusively to the eMBB traffic. That is, orthogonal slicing means that the eMBB and MTC devices share the channel in a TDMA manner. On the other hand, the whole timeslot is allocated to eMBB and mMTC under the non-orthogonal slicing, thus there is an overlap of the traffic from both services during the whole timeslot.

\par As in \cite{popovski2018}, we assume a standard scheduled transmission phase for the eMBB traffic, where the scheduling of the eMBB device has been solved prior to the considered timeslot. The frequency channel $f$ is assumed to be within the time- and frequency-coherence interval, so that the wireless channel coefficients are constant within each timeslot and also fade independently across the timeslots. The channel gains of the eMBB and MTC devices as seen by the receive antenna $l$, which we denote by $g_{i,l}$ with $i \in \left\{B,M\right\}$, are independent and identically distributed (i.i.d.), and follow a zero-mean complex Gaussian distribution with variance $\Gamma_i$, i.e., Rayleigh fading. This is, $g_{i,l} \sim \mathcal{CN}(0,\Gamma_i)$, where $\Gamma_i$ is the average channel gain.

\par Let us denote by $\textbf{
g}_i=[g_{i,1},\ldots,g_{i,L}]^T$ the vector of the wireless channel gains from the eMBB or MTC device to the BS as seen by the $L$ receive antennas. In the case of interference-free transmissions, the received Signal-to-Noise Ratio (SNR) obtained after applying MRC is given by
\begin{equation}
    \gamma_i=\Vert\textbf{g}_i\Vert^2.
\end{equation}

\par In our model, we assume that the average transmit power of all devices is normalized to one, while the differences in the actual transmit power of the devices and in the path loss are accounted for through the average channel gains $\Gamma_i$. Moreover, we also consider that the noise power at the receiver is normalized to one, such that the received power equals the SNR for all the devices.

\par No Channel-State Information (CSI) is assumed at the MTC devices, whereas the eMBB device and BS are assumed to have perfect CSI, as in \cite{popovski2018}. As a consequence, the eMBB device can adapt its transmit power according to the channel conditions such that its achievable data rate equals a predefined value. Since the MTC devices operate without CSI, they all transmit with the same fixed data rate.

\par The outage probabilities of the eMBB and mMTC services are denoted as $\Pr(E_B)$ and $\Pr(E_M)$, respectively, and must satisfy the reliability requirements $\Pr(E_B) \leq \epsilon_B$ and $\Pr(E_M) \leq \epsilon_M$.

\par In the following subsections, we present the performance analysis of the eMBB and mMTC services when operating in an orthogonal fashion, by extending the results from \cite{popovski2018} to a scenario where the receiver is provided with multiple antennas operating under MRC.

\subsection{Performance Analysis of eMBB}

\par The eMBB device adapts its transmit power $P_B(\gamma_B)$ according to the instantaneous channel gains such that the received SNR always equals a predefined value. Following~\cite{popovski2018},  the objective of the eMBB device is to transmit at the largest rate $r_B$ that is compatible with the outage probability requirement $\epsilon_B$ under a long-term average power constraint. This can be formulated as the following optimization problem
\begin{equation}
    \label{optimal rb}
	\begin{aligned}
		\text{maximize }&r_B \\
		\text{subject to }&\text{Pr}\left\{\text{log}_2[1 + P_B(\gamma_B)\gamma_B]\leq r_B\right\} \leq \epsilon_B \\
		\text{and }&\mathbb{E}[P_B(\gamma_B)] = 1.
	\end{aligned}	
\end{equation}
The optimal solution to this problem is given by the truncated power inversion scheme: the eMBB device chooses a transmit power that is inversely proportional to the received SNR $\gamma_B$ if the latter is above a given threshold $\gamma_B^{\min}$, while it refrains from transmitting otherwise \cite{popovski2018}. Thus, the activation probability of the eMBB device can be written as \cite[Eq. 7.17]{goldsmith}
\begin{align}
    a_B &= \Pr\left\{\gamma_B\geq \gamma_B^{\text{min}}\right\} \notag\\
    &= \exp\left(-\dfrac{\gamma_B^{\text{min}}}{\Gamma_B}\right)\sum_{l=1}^{L}\dfrac{(\gamma_B^{\min}/\Gamma_B)^{l-1}}{(l-1)!} \notag\\
    &= \dfrac{\Gamma(L,\gamma_B^{\min}/\Gamma_B)}{(L-1)!} \label{a_B},
\end{align}
where $\Gamma(a,z)= \int_{z}^{\infty} t^{a-1}e^{-t}\mathrm{d}t$ is the upper incomplete gamma function.

\par In the absence of interference from the mMTC traffic, the only source of outage for an eMBB transmission is the non-transmission event because of extremely poor channel conditions. In this case, the outage probability of the eMBB device can be written as 
\begin{equation}
\label{Pr_Eb}
    \Pr(E_B) = \Pr\left\{\gamma_B < \gamma_B^{\min}\right\} =
    1 - a_B,
\end{equation}
where $a_B$ is given by (\ref{a_B}).

\par Imposing the reliability requirement $\Pr(E_B) = \epsilon_B$ on (\ref{Pr_Eb}), we obtain the threshold SNR as
\begin{equation}
\label{Gb_min}
	\gamma_B^{\min}=\Gamma_B\gamma^{-1}(L,\epsilon_B(L-1)!).
\end{equation}

\par Based on the truncated power inversion scheme, the instantaneous power $P_B(\gamma_B)$ chosen as a function of the received SNR $\gamma_B$ is
\begin{equation}
	P_B(\gamma_B) = 	\begin{cases}
						\dfrac{\gamma_B^{\text{tar}}}{\gamma_B} &\text{if } \gamma_B \geq \gamma_B^{\min} \\
						0 &\text{if } \gamma_B < \gamma_B^{\min}
					\end{cases},
\end{equation}
where $\gamma_B^{\text{tar}}$ is the target SNR, which is obtained by imposing the average power constraint as \cite{popovski2018}
\begin{align}
    \mathbb{E}[P_B(\gamma_B)] &= \int_{\gamma_B^{\min}}^{\infty} \dfrac{\gamma^{L-1}e^{-\gamma/\Gamma_B}}{\Gamma_B^L(L-1)!}P_B(\gamma)d\gamma \notag\\
   &= \dfrac{\gamma_B^{\text{tar}}}{\Gamma_B(L-1)!} \Gamma \left(L-1,\dfrac{\gamma_B^{\min}}{\Gamma_B}\right) = 1.
\end{align}

\par This implies that the target SNR is
\begin{equation}
    \label{Gb_tar}
    \gamma_B^{\text{tar}} = \dfrac{\Gamma_B(L-1)!}{\Gamma \left(L-1,\tfrac{\gamma_B^{\min}}{\Gamma_B}\right)}.
\end{equation}

\par Finally, the outage rate achieved by the eMBB device is \cite{popovski2018}
\begin{equation}
    \label{rb_orth}
	r_B^{\text{out}} = \log_2(1 + \gamma_B^{\text{tar}}).
\end{equation}

\par In the discussion above, it can be noted that the outage probability of the eMBB device is uniquely determined by the imposed reliability requirement $\epsilon_B$.

\subsection{Performance Analysis of the mMTC}
\label{Performance Analysis of the mMTC Users}

\par We assume that $M$ MTC devices are connected to the BS.
Following \cite{liu2016} and considering the absence of interference from the eMBB traffic, the $L \times 1$ baseband received vector at the BS is given by
\begin{equation}
    \textbf{y}=\sqrt{P_M}\textbf{G}_M\textbf{x}_M+ \textbf{n},
\end{equation}
where $\textbf{G}_M \in \mathbb{C}^{L \times M}$ is the matrix of channel gains between the MTC devices and the BS, $\sqrt{P_M}\textbf{x}_M \in \mathbb{C}^{M\times1}$ is the vector of symbols transmitted by the MTC devices, and $\textbf{n} \in \mathbb{C}^{L\times1}$ is the vector of Additive White Gaussian Noise (AWGN) samples with zero mean and unit variance. The $m$-th element of $\textbf{x}_M$, $x_m$, is zero if the $m$-th MTC device is not active in the timeslot.

\par By exploiting the perfect CSI, the BS utilizes a MRC-SIC iterative receiver to decode the signals from the multiple MTC devices that arrive in the same timeslot. Following \cite{liu2016}, the received signal vector after the MRC processing is given by
\begin{equation}
    \hat{\textbf{x}}=\textbf{G}_M^H\textbf{y}=\sqrt{P_M}\textbf{G}_M^H\textbf{G}_M\textbf{x}_M + \textbf{G}_M^H\textbf{n},
\end{equation}
where $\hat{\textbf{x}}\in\mathbb{C}^{M\times1}$ and the superscript $H$ indicates the conjugate transpose of the matrix $\textbf{G}_M$.

\par Let $\hat{x}_m$ denote the $m$-th element of the vector $\hat{\textbf{x}}$, which corresponds to the signal transmitted by the $m$-th MTC device. As in \cite{liu2016}, we have
\begin{equation}
    \label{r_M}
    \hat{x}_m=\sqrt{P_M}\textbf{g}_m^H\textbf{g}_mx_m+\sqrt{P_M}\textbf{g}_m^H\sum_{m'\neq m}^{M}\textbf{g}_{m'}x_{m'}+\textbf{g}_{m}^H\textbf{n},
\end{equation}
where $\textbf{g}_m \in \mathbb{C}^{L\times1}$ denotes the $m$-th column of the matrix $\textbf{G}_M$. The first term in (\ref{r_M}) represents the signal transmitted by the $m$-th MTC device, while the remaining terms represent the interfering signals from other MTC devices and the noise.

\par Since  MTC devices are operate without CSI, they all transmit with the same power and have the same target data rate $r_M$. During the MRC-SIC decoding, first the BS detects the strongest device among the active MTC devices, decodes its signal, subtracts its interference from the received signal, proceeds to the second strongest MTC device, and so on. The decoding procedure ends when the decoding of one MTC device fails or after all the active MTC devices are correctly decoded.

\par The SIC decoding ordering is defined in the descending order of received SNRs of the active MTC devices. Let us denote a SIC decoding ordering $\left\{1,\ldots,M\right\}$, such that $$\textbf{g}_1^H\textbf{g}_1 \geq \textbf{g}_2^H\textbf{g}_2 \geq \ldots \geq \textbf{g}_{M}^H\textbf{g}_{M}.$$ The Signal-to-Interference-plus-Noise Ratio (SINR) while decoding the signal from the $m$-th MTC device, and assuming that the MTC devices with indexes $\left\{1,\ldots,m-1\right\}$ have already been correctly decoded, reads
\begin{equation}
    \sigma_m=\dfrac{P_M\Vert\textbf{g}_m\Vert^4}{P_M\sum\limits_{m'=m+1}^{M} |\textbf{g}_m^H\textbf{g}_{m'}|^2+\Vert \textbf{g}_m\Vert^2}.
\end{equation}
Then, the $m$-th MTC device is correctly decoded if the inequality $\log_2(1+\sigma_m)\geq r_M$ holds.

\section{Slicing between eMBB and mMTC}
\label{Slicing between eMBB and mMTC}

\par In this section we discuss the two different slicing schemes that allow the eMBB and mMTC services to coexist in the same RAN. For both schemes we characterize the pair of maximum achievable data rates $(r_B,r_M)$ given the reliability requirements $\epsilon_B$ and $\epsilon_M$ for eMBB and mMTC, respectively.

\subsection{Orthogonal Slicing between eMBB and mMTC}

\par Under the orthogonal slicing, the eMBB device and the MTC devices use the radio resource in a time-sharing manner. Let $\alpha \in [0,1]$ and $1-\alpha$ denote the fraction of time in which the frequency channel is allocated to the eMBB traffic and to the mMTC traffic, respectively. For a given time-sharing factor $\alpha$, the eMBB data rate is \cite{popovski2018}
\begin{equation}
\label{r_B}
    r_B=\alpha r_B^{\text{out}},
\end{equation}
where $r_B^{\text{out}}$ is given by (\ref{rb_orth}). Similarly, the mMTC data rate is
\begin{equation}
    r_M=(1-\alpha)r_M^{\text{out}},
\end{equation}
where $r_M^{\text{out}}$ is the maximum achievable mMTC data rate in the absence of interference from the eMBB traffic.

\par To characterize the performance of the orthogonal slicing, for each value of $\alpha$, we set an eMBB data rate according to (\ref{r_B}). Then we compute the maximum achievable mMTC data rate $r_M$ for which the reliability requirements $\epsilon_B$ and $\epsilon_M$ are met. 

\subsection{Non-Orthogonal Slicing between eMBB and mMTC}

\par Under the non-orthogonal slicing, the eMBB and mMTC traffics overlap in the radio resource. The received signal vector at the BS is then
\begin{equation}
    \textbf{y}=\textbf{G}\textbf{x}+\textbf{n},
\end{equation}
where
\begin{equation}
    \textbf{G}=[\textbf{g}_{m,1}\;\textbf{g}_{m,2}\;\ldots\;\textbf{g}_{m,M}\;\textbf{g}_B]
\end{equation}
is a matrix containing the channel gains between all the devices and the BS, $\textbf{G} \in \mathbb{C}^{L\times(M+1)}$,
\begin{equation}
    \textbf{x}=[\sqrt{P_M}[x_{m,1}\;x_{m,2}\;\ldots\;x_{m,M}]\;\sqrt{P_B}x_B]^T
\end{equation}
is the complex vector containing the transmitted symbols from the MTC devices and from the eMBB device, and $\textbf{n}\in\mathbb{C}^{L\times1}$ is the vector containing the noise samples. As in the orthogonal case, the received signal vector after the MRC at the BS is
\begin{equation}
    \hat{\textbf{x}}=\textbf{G}^H\textbf{y}=\textbf{G}^H\textbf{G}\textbf{x}+\textbf{G}^H\textbf{n}.
\end{equation}

\par Let us denote $\hat{x}_m$ the element of the vector $\hat{\textbf{x}}$ corresponding to the signal of the $m$-th MTC device, and $\hat{x}_B$ the element of the vector $\hat{\textbf{x}}$ corresponding to the signal of the eMBB device. They read
\begin{equation}
\begin{split}
    \label{r_M_non}
    \hat{x}_m=\sqrt{P_M}\textbf{g}_m^H\textbf{g}_mx_m+\sqrt{P_M}\textbf{g}_m^H\sum_{m'\neq m}^{M}\textbf{g}_{m'}x_{m'}+\\
    \sqrt{P_B}\textbf{g}_m^H\textbf{g}_Bx_B
    +\textbf{g}_{m}^H\textbf{n},
\end{split}
\end{equation}
\begin{equation}
    \label{r_B}
    \hat{x}_B=\sqrt{P_B}\textbf{g}_B^H\textbf{g}_Bx_B+\sqrt{P_M}\textbf{g}_B^H\sum_{m=1}^{M}\textbf{g}_{m}x_{m}+\textbf{g}_{B}^H\textbf{n}.
\end{equation}
Assuming the SIC decoding ordering $\left\{1,\ldots,M\right\}$ as in the orthogonal case, the SINR of the  $m$-th mMTC in the presence of the eMBB interference reads
\begin{equation}
    \sigma_m=\dfrac{P_M\Vert\textbf{g}_m\Vert^4}{P_M\sum\limits_{m'=m+1}^{M} |\textbf{g}_m^H\textbf{g}_{m'}|^2 + P_B|\textbf{g}_m^H\textbf{g}_B|^2 + \Vert \textbf{g}_m \Vert^2}.
\end{equation}
As in the orthogonal case, the mMTC is correctly decoded if the inequality $\log_2(1+\sigma_m)\geq r_M$ holds.

\par After the correct decoding of the $m$-th MTC device, the BS attempts to decode the eMBB device if it has not been decoded yet. Then the SINR of the eMBB device reads
\begin{equation}
    \sigma_B=\dfrac{P_B\Vert\textbf{g}_B\Vert^4}{P_M\sum\limits_{m'=m+1}^{M} |\textbf{g}_B^H\textbf{g}_{m'}|^2 + \Vert\textbf{g}_B\Vert^2}.
\end{equation}
For a given data rate $r_B$, the eMBB device is correctly decoded if the inequality $\log_2(1+\sigma_B)\geq r_B$ holds.

\par Under the orthogonal slicing, the eMBB device adopts a fixed target SNR $\gamma_B^{\text{tar}}$ that satisfies the power constraint $\mathbb{E}\left\{P_B\right\}=1$. On the contrary, aiming to minimize the interference that the eMBB traffic causes to the mMTC traffic, under the non-orthogonal slicing we allow the eMBB device to adopt lower values for the target SNR, which yields the inequality \cite{popovski2018}
\begin{equation}
    \gamma_{B}^{\text{tar}} \leq \dfrac{\Gamma_B(L-1)!}{\Gamma \left(L-1,\tfrac{\gamma_{B}^{\min}}{\Gamma_B}\right)}.
\end{equation}
Consequently, we have $\mathbb{E}\left\{P_B\right\}\leq1$. Nevertheless, this condition is acceptable when the eMBB device transmits with a data rate $r_B\leq r_B^{\text{out}}$.

\par Differently from the orthogonal slicing, the error probability for eMBB has two components in the non-orthogonal case: the probability of the eMBB device does not transmit due to insufficient SNR; and the probability of the eMBB device transmits because it has sufficient SNR, but a decoding error occurs due to the interference from the mMTC traffic. In order to satisfy the same reliability requirement from the orthogonal case, we must allow the activation probability of the eMBB device to be higher, which yields $a_B>1-\epsilon_B$. If we adopt, for example, $\epsilon_B=10^{-3}$, then $a_B>0.999$. For the computation of the maximum achievable mMTC data rate, we conservatively assume that the eMBB interference is always present, that is, $a_B=1$, such that the error probability for eMBB is just the decoding error probability, that is,
\begin{equation}
    \Pr(E_B)=\Pr\left\{\log_2(1+\sigma_B)<r_B\right\}.
\end{equation}

\par The SIC decoding procedure for the non-orthogonal slicing runs as follows. Initially, all the MTC devices suffer with the interference from eMBB traffic. First the BS attempts to decode the strongest MTC device. If the decoding succeeds, the signal from the decoded device is subtracted from the received signal, BS attempts do decode the second strongest MTC device, and so on. If the decoding of a MTC device fails, the BS tries to decode the signal from the eMBB device. If its signal is correctly decoded, the interference from eMBB is subtracted from the received signal, then the BS returns to the decoding of the MTC devices, and the procedure continues as described in Section \ref{Performance Analysis of the mMTC Users}. Otherwise, if the decoding of the eMBB fails, the SIC decoding ends. The other condition that terminates the SIC decoding procedure is when all the MTC devices are correctly decoded, and so the last step is just the decoding of the eMBB signal without the interference from the mMTC traffic. It is important to note that the step when the eMBB device is decoded is random.

\par The performance characterization of the non-orthogonal slicing is a two dimensional numerical search: first we set an eMBB data rate $r_B \in [0,r_B^{\text{out}}]$ and then compute the maximum number achievable mMTC data rate $r_M$ that is adopted for all the MTC devices connected to the BS while still satisfying the reliability requirements $\epsilon_B$ and $\epsilon_M$; during this computation, we seek for the minimum value of $\gamma_B^{\text{tar}}$ that can be adopted by the eMBB device. 

\section{Numerical Results}
\label{numerical illustration}

\par In this section we present Monte Carlo simulation results to illustrate the performance of both orthogonal and non-orthogonal slicing of radio resources between eMBB and mMTC. As the general parameters, we set the reliability requirements $\epsilon_B=10^{-3}$ and $\epsilon_M=10^{-1}$ for eMBB and mMTC, respectively, and also the average channel gains $\Gamma_B=20$ dB and $\Gamma_M=5$ dB for eMBB and mMTC, respectively.

\par
\begin{figure}
    \centering
    \includegraphics[scale=0.5]{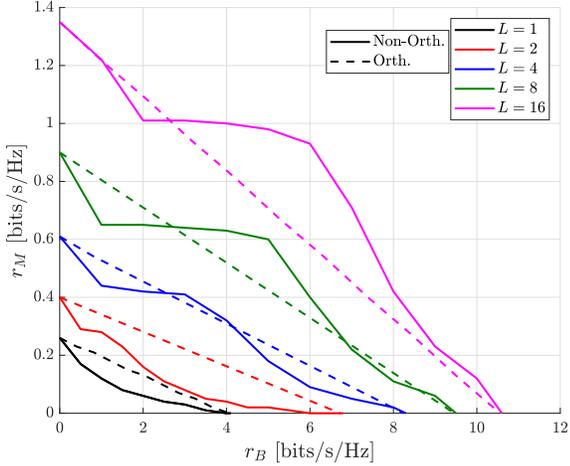}
    \caption{eMBB data rate $r_B$ versus mMTC data rate $r_M$ for the orthogonal and non-orthogonal slicing, considering different numbers of receive antennas and for $\Gamma_B=20$ dB, $\Gamma_m=5$~dB, $\epsilon_B=10^{-3}$, $\epsilon_m=10^{-1}$ and $M=10$.}
    \label{results1}
\end{figure}

\begin{figure}
    \centering
    \includegraphics[scale=0.5]{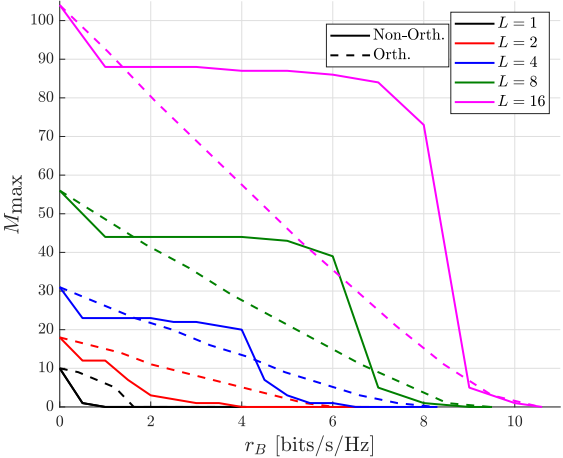}
    \caption{eMBB data rate $r_B$ versus the maximum number of connected MTC devices $M_{\text{max}}$ for the orthogonal and non-orthogonal slicing, considering different numbers of receive antennas and for $\Gamma_B=20$ dB, $\Gamma_m=5$~dB, $\epsilon_B=10^{-3}$, $\epsilon_m=10^{-1}$ and $r_M=0.25$ bits/s/Hz.}
    \label{results5}
\end{figure}

 In Fig. \ref{results1} we plot the pairs of achievable data rates $(r_M,r_B)$ for a given number $M=10$ of MTC devices connected to the BS, and for different values of receive antennas $L \in \left\{1,2,4,8,16\right\}$. 
Conversely, in Fig. \ref{results5} we plot the pairs $(M_{\text{max}},r_M)$ of maximum number of connected MTC devices versus the eMBB data rate for a given mMTC data rate $r_M=0.25$ bits/s/Hz, and also for $L \in \left\{1,2,4,8,16\right\}$. The dashed curves correspond to the orthogonal slicing, while the continuous curves correspond to the non-orthogonal slicing.

\par As observed in both figures, increasing $L$ always increase the performance of the system for both slicing schemes. At each SIC decoding step the MRC receiver projects the received signal into the direction of the signal of interest. As we increase the number of antennas, we increase the diversity gain, so the components of the received signal are boosted into the direction of the signal of interest and attenuated on other directions, which represents a boost on the available SINR at each SIC decoding step.

\par Meanwhile, we also observe that as we increase the number of receive antennas, the non-orthogonal slicing outperforms more easily the orthogonal slicing both in terms of the maximum achievable mMTC data rate $r_M$ and of the maximum number of connected MTC devices $M_{\text{max}}$. For $L=1$ and $L=2$, the orthogonal slicing outperforms the non-orthogonal slicing for the whole range of $r_B$. However, as we increase the number of receive antennas to $L\geq4$, the non-orthogonal slicing becomes increasingly more advantageous over the orthogonal slicing for the intermediate and higher values of $r_B$ because it allows us to achieve pairs $(r_M,r_B)$ and $(M_{\text{max}},r_B)$ that are not possible to achieve with the orthogonal slicing.

\par The curves for the orthogonal slicing are straight lines because $r_M$ and $r_B$ in Fig. \ref{results1} and $M_\text{max}$ and $r_B$ in Fig. \ref{results5} are linearly scaled according to the fraction of the timeslot $\alpha$ that is allocated for each service. On the other hand, the curves for the non-orthogonal cases present a non-linear shape because it is defined according to the level of interference that eMBB causes to mMTC. Starting from $r_B=0$, there is no interference from the eMBB traffic, so the mMTC performance for both slicing schemes are the same. Then, when $r_B>0$, there is an abrupt reduction in the performance of mMTC because of the presence of interference from eMBB traffic. For the lowest values of $r_B$, the interference that the eMBB traffic causes to the mMTC traffic is minimal, so almost all the MTC devices are correctly decoded before the decoding of the eMBB device. As we increase $r_B$ to the intermediate values, we also increase the interference from eMBB because the eMBB device has to adopt higher values for the target SNR to meet the target data rate. In this regime, the eMBB device starts to be decoded before some of the MTC devices have been decoded. As a consequence, after the correct decoding of eMBB, some of the MTC devices do not suffer with the interference anymore, and the decrease in the performance of mMTC is very smooth. Finally, for the higher values of $r_B$, the eMBB device has to adopt higher values for its target SNR, which causes more interference to the mMTC traffic. In this regime, the performance of eMBB is also severely limited by the interference from the mMTC traffic.

\par It is important to note that even tough increasing $L$ substantially increases the performance of mMTC, in term of both achievable data rates and number of connected devices, it also increases the receiver complexity. Moreover, NOMA of a massive number of MTC devices also increases the receiver complexity and yields higher processing delay times. These aspects must be taken into account when implementing the network slicing of radio resources in practical situations.


\section{Conclusions}
\label{conclusions}

\par We studied the coexistence of eMBB and mMTC in the uplink of the same RAN, enabled by two different network slicing schemes. Both services share a radio resource that is composed of a single frequency channel and a single timeslot. Under the orthogonal slicing, a fraction of the timeslot is allocated exclusively for mMTC, while the remaining of the timeslot is allocated exclusively for eMBB. On the other hand, under the non-orthogonal slicing, the traffic from both services overlap during the whole duration of the time slot. In both schemes, the massive connectivity required by mMTC applications is achieved through the use of NOMA with SIC decoding. The use of multiple receive antennas mitigates the imperfections of the wireless channel and guarantees the spectral efficiency of both services. We set the reliability requirements and then evaluated the pairs of maximum achievable data rates through Monte Carlo simulations. Our simulation results showed that, the more we increase the number of receive antennas, the more advantageous the non-orthogonal slicing becomes over the orthogonal slicing in term of both the achievable mMTC data rates for a given number of connected devices, and the number of connected MTC devices for a given mMTC data rate. Finally, although the spatial receive diversity increases substantially the performance of the system, it also increases the receiver complexity. Moreover, NOMA of a massive number of devices is also a complex task and yields higher processing delay times. Such aspects must be considered in practical implementations.

\section*{Acknowledgment}

\par This research has been financially supported by Academy of Finland, 6Genesis Flagship (grant no 318927), FIREMAN (no 326301) and Aka Prof (no 307492).

\bibliographystyle{./bibliography/IEEEtran}
\bibliography{./bibliography/references}

\end{document}